\documentclass[twocolumns,amsmath,amssymb,prl]{revtex4}
\usepackage{graphicx}% Include figure files
\usepackage{dcolumn}% Align table columns on decimal point
\usepackage{bm}% bold math

\begin{document}

%\preprint{APS/123-QED}

\title{Electrical and thermal spin accumulation in germanium}% Force line breaks with \\

\author{A. Jain$^{1}$, C. Vergnaud$^{1}$, J. Peiro$^{2}$, J. C. Le Breton$^{2}$, E. Prestat$^{1}$, , L. Louahadj$^{1}$, C. Portemont$^{3}$, C. Ducruet$^{3}$, V. Baltz$^{4}$, A. Marty$^{1}$, A. Barski$^{1}$, P. Bayle-Guillemaud$^{1}$, L. Vila$^{1}$, J.-P. Attan\'e$^{1}$, E. Augendre$^{5}$, H. Jaffr\`es$^{2}$, J.-M. George$^{2}$ and M. Jamet$^{1*}$}
\affiliation{$^{1}$INAC/SP2M, CEA-UJF, F-38054 Grenoble, France\\
$^{2}$UMP CNRS-Thal\`es, CNRS, F-91767 Palaiseau, France\\
$^{3}$CROCUS-Technology, F-38025 Grenoble, France\\
$^{4}$INAC/Spintec, CEA-CNRS-UJF-INPG, F-38054 Grenoble, France\\
$^{5}$LETI, CEA, Minatec Campus, F-38054 Grenoble, France\\
$^{6}$INAC/SCIB, CEA-UJF, F-38054 Grenoble, France}%

\date{\today}% It is always \today, today,
             
\begin{abstract}
In this letter, we first show electrical spin injection in the germanium conduction band at room temperature and modulate the spin signal by applying a gate voltage to the channel. The corresponding signal modulation agrees well with the predictions of spin diffusion models. Then by setting a temperature gradient between germanium and the ferromagnet, we create a thermal spin accumulation in germanium without any tunnel charge current. We show that temperature gradients yield larger spin accumulations than pure electrical spin injection but, due to competing microscopic effects, the thermal spin accumulation in germanium remains surprisingly almost unchanged under the application of a gate voltage to the channel.
\end{abstract}

%\pacs{75.50.Pp, 75.75.-c, 75.30.Gw, 61.46.-w, 76.30.-v}

\maketitle

The development of semiconductor (SC) spintronics requires nowadays the creation of a spin polarized carrier population in the SC conduction band up to room temperature\cite{Awschalom2007,Zutic2004}. The SC should further be compatible with microelectronics technology \textit{i.e.} made of silicon or germanium\cite{Zutic2011,Dery2007,Dery2011}. For that purpose, several methods based on spin-dependent phenomena have been explored in the case of pure silicon and pure germanium: direct electrical spin injection from a ferromagnetic metal through a tunnel barrier\cite{Dash2009,Jonker2007,Appelbaum2007,Li2011,Suzuki2011,Zhou2011,Jain2011,Jeon2011,Saito2011,Jain2012} charge-to-spin current conversion using spin-orbit based effects like spin Hall effect\cite{Jain2012,Ando2012} and spin-dependent thermoelectric effects such as the Seebeck spin tunneling recently observed in silicon\cite{Lebreton2011,Jansen2012}. In this letter we have successfully achieved electrical and thermal spin injection in the conduction band of germanium at room temperature using the same three-terminal device. We have further compared their respective efficiency and modulated the spin signals by applying a back gate voltage. In this paper, all the measurements have been performed at room temperature. The gate voltage dependence of the electrical spin signal clearly follows the predictions of standard spin diffusion models whereas the thermal spin signal remains surprisingly almost unchanged under the application of a gate voltage probably due to compensating effects.\\

%\begin{figure}[h!]
%\begin{center}
%\includegraphics[width=0.46\textwidth]{Fig1a} \\
%\includegraphics[width=0.46\textwidth]{Fig1b} \\
%\includegraphics[width=0.4\textwidth]{Fig1c} 
%\end{center}
%\caption{(color online) Three-terminal geometry used for (a) electrical spin injection (contact geometry A) and (b) thermal spin accumulation (contact geometry B) in germanium. A back gate voltage $V_{G}$ can be applied to the germanium channel. (c) Cross section TEM image of the ferromagnet/tunnel barrier electrode to perform spin injection/detection in germanium. Even after annealing, the Ta/CoFeB/MgO stack appears as almost amorphous.}
%\label{fig1}
%\end{figure}

The multi-terminal device we used for electrical and thermal spin injection is shown in Fig. 1(a) and 1(b). The full stack Ta(5 nm)/Co$_{60}$Fe$_{20}$B$_{20}$(5 nm)/MgO(3 nm) has been grown by sputtering and annealed on Germanium-On-Insulator (GOI) wafers\cite{Jain2012} (Fig. 1(c)). GOI substrates were fabricated using the SmartCut$^{TM}$ process and Ge epitaxial wafers\cite{Deguet2005}. The transferred 40 nm-thick Ge film was $n$-type doped in two steps: a first step (phosphorus, 3$\times$10$^{13}$ cm$^{-2}$, 40 keV, annealed for 1h at 550$^{\circ}$C) that provided uniform doping in the range of 10$^{18}$ cm$^{-3}$, and a second step (phosphorus, 2$\times$10$^{14}$ cm$^{-2}$, 3 keV, annealed for 10 s at 550$^{\circ}$C) that increased surface $n$+ doping to the vicinity of 10$^{19}$ cm$^{-3}$ to make the Schottky barrier transparent. The thickness of the $n$+ doped layer is estimated to be 10 nm. The surface of the GOI was finally capped with amorphous SiO$_{2}$ to prevent surface oxidation, this capping layer is removed by using hydrofluoric acid prior to the growth of Ta/CoFeB/MgO. GOI wafers are further made of a Si $p$+ degenerate substrate and 100 nm-thick SiO$_{2}$ layer (BOX) to apply a back gate voltage to the Ge film and modulate spin signals. Finally conventional optical lithography was used to define the germanium channel and three contacts made of a tunnel spin injector in between two ohmic contacts made of Au(250 nm)/Ti(10 nm). Soft argon etching is used to remove the 10 nm-thick $n$+-doped germanium layer.

%\begin{figure}[h!]
%\begin{center}
%\includegraphics[width=0.23\textwidth]{Fig2a}
%\includegraphics[width=0.23\textwidth]{Fig2b} \\
%\includegraphics[width=0.23\textwidth]{Fig2c}
%\includegraphics[width=0.23\textwidth]{Fig2d}
%\end{center}
%\caption{(color online) (a), (b) Gate voltage dependence of the tunnel junction and channel resistance respectively. (c) Spin signal recorded with the applied field in-plane ($V_{//}$) and out-of-plane ($V_{\bot}$) for two different gate voltages 0 V and -50 V. Solid lines are lorentzian fits. (d) Evolution with the gate voltage of the channel resistance, tunnel junction resistance (measured at a bias current of 100 $\mu$A) and the spin signal in \%.}
%\label{fig2}
%\end{figure} 

$I(V)$ curves have been recorded using the contact geometry A (Fig. 2(a)) and B (Fig. 2(b)) as a function of the gate voltage $V_{G}$. Geometry A probes the tunnel junction resistance which exhibits a clear non-linear symmetric behavior as expected for pure tunneling transport while geometry B probes the Ge channel resistance ($R$). As expected the tunnel junction resistance exhibits no gate voltage dependence whereas the channel resistance increases by 30 \% from $V_{G}$=0 V to $V_{G}$=-50 V. Indeed at negative gate voltage, the carrier density is lowered in the $n$-Ge channel and its resistivity is enhanced. For positive gate voltage, the channel resistance remains constant. We have then investigated the effect of this gate voltage on the spin signal. Electrical spin injection/detection measurements have been performed at room temperature using the non-local contact geometry A (Fig. 1(a)). We have recently shown that spin injection takes place in the $n$-Ge conduction band at this temperature\cite{Jain2012}. In Fig. 2(c), the magnetic field was applied out-of-plane along $z$ to obtain Hanle curves ($V_{\bot}$: spin precession around the applied field) and in-plane along $x$ to obtain inverted-Hanle curves ($V_{//}$: to suppress spin precession around interface random fields)\cite{Dash2011}. In that case, the total spin signal scaling with the full spin accumulation is given by: $V_{S}$=$V_{\bot}$+$V_{//}$ and the spin resistance-area product by: $R_{S}A$=$(V_{S}/I).A$ where $I$ is the applied current and $A$ the ferromagnetic electrode area. The total spin signal increases with negative gate voltage and the results are summarized in Fig. 2(d). For $V_{G}$=-50 V, we obtain $\Delta V_{S}/V_{S}$=$(V_{S}(-50V)-V_{S}(0V))$/$V_{S}(0V)\approx \Delta R/R\approx$30 \%. To be more quantitative, in the case of spin injection in the Ge conduction band and in the frame of the diffusive regime model\cite{Fert2001}, the spin resistance-area product is given by: $R_{S}A$=$(V_{S}/I).A$=$(TSP\times l_{sf}^{cb})^{2}\times (\rho /t_{Ge})$ where $TSP$ is the tunnel spin polarization, $l_{sf}^{cb}$ the spin diffusion length in the germanium conduction band and $\rho$ (resp. $t_{Ge}$) the germanium resistivity (resp. thickness). Hence if we assume that $TSP$ and $l_{sf}^{cb}$ remain constant under the application of an electric field, $V_{S}$ scales as $(\rho /t_{Ge})$ which is proportional to the channel resistance $R$. We then expect $\Delta V_{S}/V_{S}$ to scale with $\Delta R/R$ in the event that spin polarized carriers are injected in the Ge conduction band. We indeed observe the spin signal increasing with the negative gate voltage down to -50 V.

%\begin{figure}[h!]
%\begin{center}
%\includegraphics[width=0.4\textwidth]{Fig3a} \\
%\includegraphics[width=0.4\textwidth]{Fig3b} \\
%\includegraphics[width=0.4\textwidth]{Fig3c}
%\end{center}
%\caption{(color online) (a) Comparison between electrically (open squares) and thermally (open circles) created spin accumulations in germanium. Solid lines are Lorentzian fits. (b) Current dependence of electrically and thermally created spin accumulations in germanium along with the corresponding linear and parabolic fits (solid lines). (c) Gate voltage dependence of the thermally created spin accumulation. Solid lines are parabolic fits and a characteristic error bar is reported on the graph.}
%\label{fig3}
%\end{figure}

The contact geometry B is used to set a temperature gradient between Ge and CoFeB by Joule heating. A charge current up to 10 mA (heating power density: 20 $\mu$W.$\mu$m$^{-3}$) is applied in the Ge channel to rise its temperature. The temperature difference between Ge and CoFeB: $\Delta T$=$T_{Ge}$-$T_{CoFeB}$ leads to a spin accumulation in Ge without any tunnel charge current provided that the tunnel spin polarization $TSP$ is energy dependent\cite{Lebreton2011,Jansen2012}. The Hanle and inverted Hanle curves as the ones shown in Fig. 3(a) are identical for both directions of the Joule heating current which means that the sign and magnitude of spin polarization induced in Ge are the same for both current directions. Moreover, in Fig. 3(b), the clear dependence of $V_{\bot}$ and $V_{//}$ on $I^{2}$ shows that spin accumulation is related to the heating power in good agreement with the tunneling spin Seebeck mechanism. In comparison, the electrical spin accumulation is linearly dependent on the applied current. In a previous work, we have shown that the spin resistance-area product $R_{S}A$ in Ge at room temperature shows no bias voltage dependence\cite{Jain2012}: $R_{S}A$=$(V_{S}/I).A$=$cte$. Hence $V_{S}$ scales linearly with the applied current $I$ in good agreement with our findings. In the whole current range, tunneling spin Seebeck is the most efficient mechanism to create spin accumulation in Ge\cite{note1}. Considering the geometry of our device, the heat mainly flows from the Ge film through the thin BOX layer down to the Si substrate. This makes our device far from being designed to create a large temperature difference between Ge and the ferromagnet: it is rather designed to apply an electric field to the Ge channel using a back gate voltage. This heat leakage through the BOX layer also explains why we have to inject so much heating power to observe thermal spin accumulation. However any quantitative estimation of the temperature difference between Ge and CoFeB is clearly out of the scope of this letter since several parameters are unknown like the interface thermal resistances, heat radiation from the sample and lateral versus vertical heat flows. \\
Finally, in Fig. 3(c), we have investigated the gate voltage dependence of thermally created spin accumulation in Ge. Surprisingly, at least in the low current regime ($<$5 mA), spin signals remain unchanged under the application of $V_{G}$=-50 V. Indeed in this regime and at constant current $I$, the Ge channel resistance $R$ and thus the heating power $RI^{2}$ increases by $\approx$30 \% up to $V_{G}$=-50 V. In the linear response regime, $V_{S}$ is expected to scale with the heating Joule power and hence should also increase with a negative gate voltage\cite{Lebreton2011}. From a microscopic point of view, the thermal spin accumulation ($V_{S}$) depends on the finite energy derivative of the $TSP$ ($\partial_{\epsilon}(TSP)$) and on the number of hot electrons promoted above the Fermi level in Ge by Joule heating. For increasing heating power, $\partial_{\epsilon}(TSP)$ remains constant whereas the number of hot electrons increases which leads to larger spin signals ($V_{S}\nearrow$). However by applying $V_{G}$=-50 V, the Fermi level in Ge is lowered to the bottom of the conduction band so that the overall number of hot electrons decreases ($V_{S}\searrow$) and both effects seem to compensate at low currents. For higher currents, some more complex mechanisms may be involved which needs further investigations.\\
In summary, we have shown both electrical and thermal spin accumulations in germanium at room temperature using the same three-terminal device. The electrical spin signal could be manipulated by applying an electric field to the Ge channel. Despite the inadequate device geometry, we could detect a thermally created spin accumulation in Ge without any tunnel charge current as a consequence of the tunneling spin Seebeck effect. Moreover, we could show that two microscopic mechanisms compensate each other when we apply a back gate voltage to the Ge channel which leads to almost no variation of the thermal spin accumulation.\\
The authors would like to acknowledge the financial support from the Nanoscience Foundation of Grenoble (RTRA project IMAGE). 

%\newpage %Just because of unusual number of tables stacked at end

%Merlin.mbs v4.21 2009-07-09.

%Merlin.mbs v4.21 2009-07-09. 
%Merlin.mbs v4.21 2009-07-09.

\newpage

\begin{figure}[h!]
%\begin{center}
%\includegraphics[width=0.46\textwidth]{Fig1a} \\
%\includegraphics[width=0.46\textwidth]{Fig1b} \\
%\includegraphics[width=0.4\textwidth]{Fig1c} 
%\end{center}
\caption{(color online) Three-terminal geometry used for (a) electrical spin injection (contact geometry A) and (b) thermal spin accumulation (contact geometry B) in germanium. A back gate voltage $V_{G}$ can be applied to the germanium channel. (c) Cross section TEM image of the ferromagnet/tunnel barrier electrode to perform spin injection/detection in germanium. Even after annealing, the Ta/CoFeB/MgO stack appears as almost amorphous.}
\label{fig1}
\end{figure}

\begin{figure}[h!]
%\begin{center}
%\includegraphics[width=0.23\textwidth]{Fig2a}
%\includegraphics[width=0.23\textwidth]{Fig2b} \\
%\includegraphics[width=0.23\textwidth]{Fig2c}
%\includegraphics[width=0.23\textwidth]{Fig2d}
%\end{center}
\caption{(color online) (a), (b) Gate voltage dependence of the tunnel junction and channel resistance respectively. (c) Spin signal recorded with the applied field in-plane ($V_{//}$) and out-of-plane ($V_{\bot}$) for two different gate voltages 0 V and -50 V. Solid lines are lorentzian fits. (d) Evolution with the gate voltage of the channel resistance, tunnel junction resistance (measured at a bias current of 100 $\mu$A) and the spin signal in \%.}
\label{fig2}
\end{figure} 

\begin{figure}[h!]
%\begin{center}
%\includegraphics[width=0.4\textwidth]{Fig3a} \\
%\includegraphics[width=0.4\textwidth]{Fig3b} \\
%\includegraphics[width=0.4\textwidth]{Fig3c}
%\end{center}
\caption{(color online) (a) Comparison between electrically (open squares) and thermally (open circles) created spin accumulations in germanium. Solid lines are Lorentzian fits. (b) Current dependence of electrically and thermally created spin accumulations in germanium along with the corresponding linear and parabolic fits (solid lines). (c) Gate voltage dependence of the thermally created spin accumulation. Solid lines are parabolic fits and a characteristic error bar is reported on the graph.}
\label{fig3}
\end{figure}

\newpage

\begin{figure}[h!]
\includegraphics[width=\textwidth]{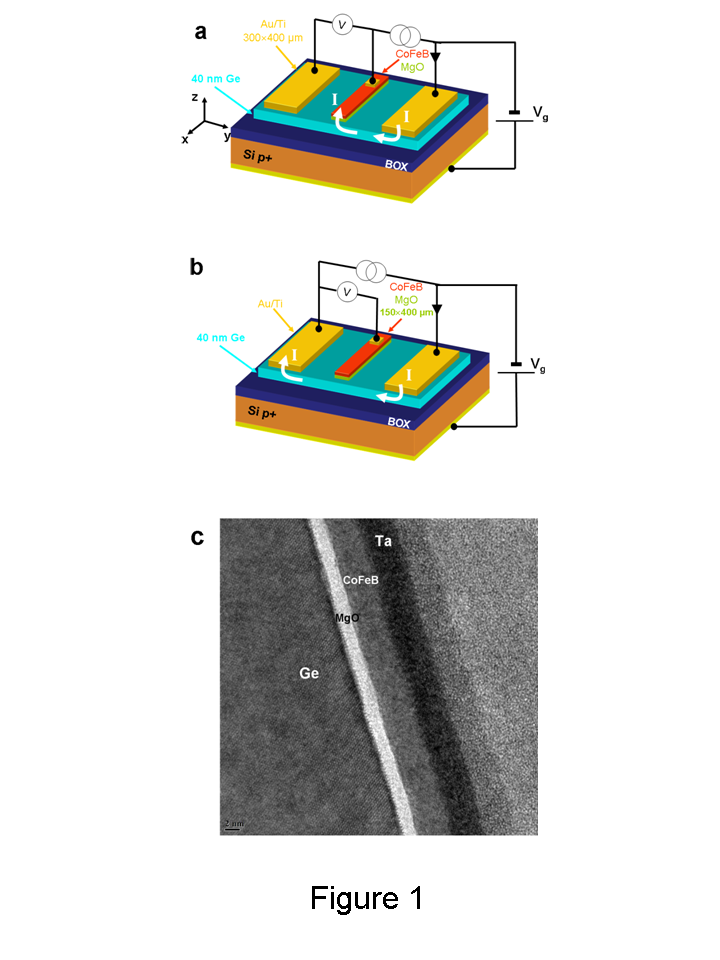}
\end{figure}

\newpage

\begin{figure}[h!]
\includegraphics[width=\textwidth]{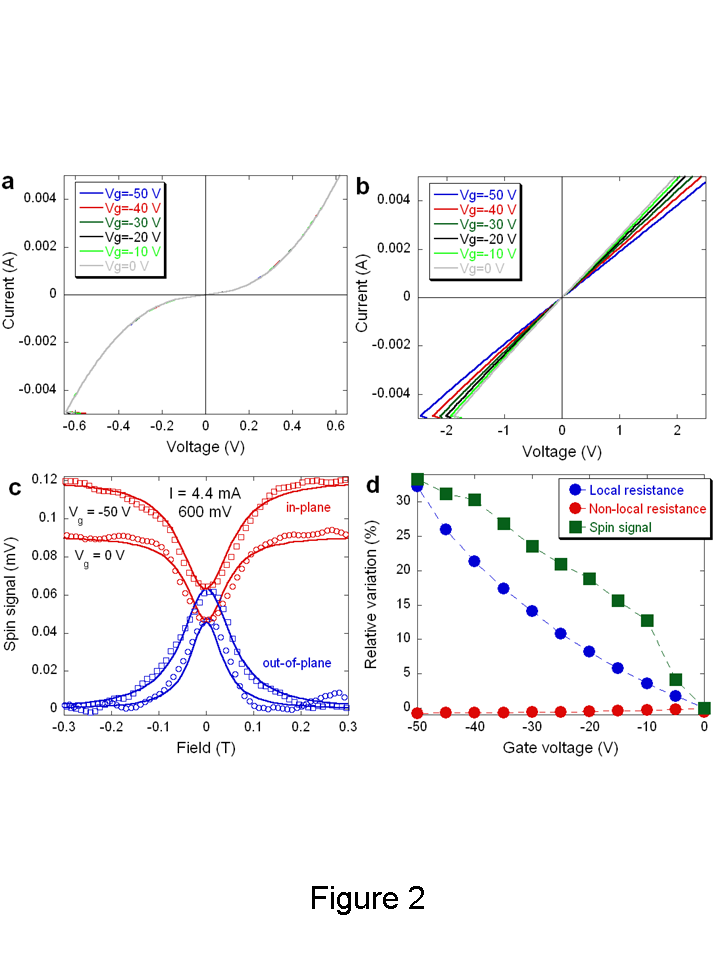}
\end{figure}

\newpage

\begin{figure}[h!]
\includegraphics[width=\textwidth]{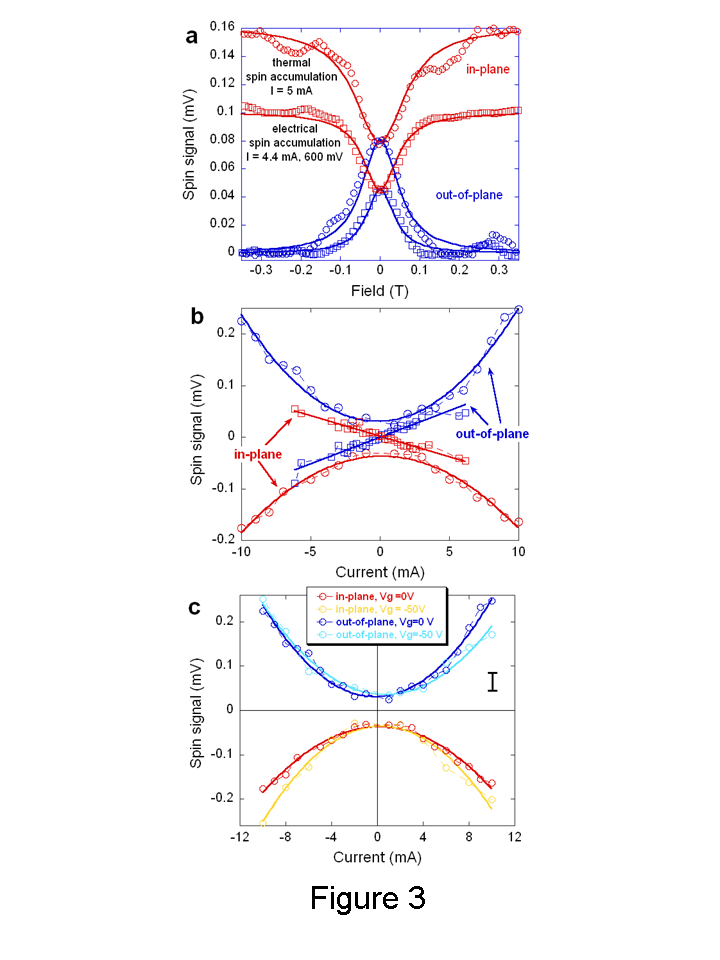}
\end{figure}


\begin{thebibliography}{}
\bibitem{note1}
The non-vanishing thermal spin signal at zero-current is an artefact of our experimental set-up. Indeed, the sample is inserted in the variable temperature insert of the cryostat and is permanently heated to keep it at room temperature. We then believe that a weak temperature difference exists between Ge and CoFeB due to this regulation that yields the non-vanishing spin signal at zero-current.
\end{thebibliography}
\end{document}